# Exploring universal patterns in human home-work commuting from mobile phone data

**Kevin S. Kung**[1,2,*]**, Kael Greco**[1,3]**, Stanislav Sobolevsky**[1] **and Carlo Ratti**[1,3]

[1] MIT Senseable City Laboratory, Cambridge, MA, USA
[2] Department of Biological Engineering, Massachusetts Institute of Technology, Cambridge, MA, USA
[3] Department of Urban Studies and Planning, Massachusetts Institute of Technology, Cambridge, MA, USA

* Corresponding author: kkung@mit.edu

Exploring universal patterns in human home-work commuting from mobile phone data

**Abstract.** Home-work commuting has always attracted significant research attention because of its impact on human mobility. One of the key assumptions in this domain of study is the universal uniformity of commute times. However, a true comparison of commute patterns has often been hindered by the intrinsic differences in data collection methods, which make observation from different countries potentially biased and unreliable. In the present work, we approach this problem through the use of mobile phone call detail records (CDRs), which offers a consistent method for investigating mobility patterns in wholly different parts of the world. We apply our analysis to a broad range of datasets, at both the country (Portugal, Ivory Coast, and Saudi Arabia), and city (Boston) scale. Additionally, we compare these results with those obtained from vehicle GPS traces in Milan. While different regions have some unique commute time characteristics, we show that the home-work time distributions and average values within a single region are indeed largely independent of commute distance or country (Portugal, Ivory Coast, and Boston)—despite substantial spatial and infrastructural differences. Furthermore, our comparative analysis demonstrates that such distance-independence holds true only if we consider multimodal commute behaviors—as consistent with previous studies. In car-only (Milan GPS traces) and car-heavy (Saudi Arabia) commute datasets, we see that commute time is indeed influenced by commute distance. Finally, we put forth a testable hypothesis and suggest ways for future work to make more accurate and generalizable statements about human commute behaviors.

**Keywords:** Mobile phone data, home-work commuting, human mobility, Marchetti's constant, constant travel budget time hypothesis

**Introduction**

With the advent of various big data initiatives and their concomitant analytics, it has become increasingly feasible to study human behavior at a massive scale . One particular avenue of research that has attracted considerable attention has been in human mobility.. Researchers have made progress in this effort using a variety of data sources such as circulating bank notes [1], taxi trip records [2, 50-51], Foursquare check-in data [3], Tweets [52], and even GPS devices [4-6].

While these different sources are promising, and lead to many similar conclusions about human mobility, they are often limited in scale (GPS traces and taxi records), limited in data resolution (bank notes), or limited in adaptation (Foursquare). In contrast, currently, mobile phone records seem to be the source that overcomes all these inherent constraints: a mobile phone is typically carried by an individual throughout the day and thus accurately tracks the mobility pattern on an individual level, and is widespread enough in terms of adaptation—even in developing countries—that it allows us to adequately sample the country-wide population (unlike taxi, Foursquare, or GPS traces). Indeed, previous studies have utilized cellphone call detail records (CDRs) to infer various characteristics of human mobility. For example, González *et al.* [7], using a European dataset, quantified the scale-free nature of human mobility at different length scales. Likewise, Song *et al.* [8] subsequently answered the more fundamental question about how predictable human mobility is from the CDR data. Simini *et al.* [9], were able to propose a universal model for human mobility based on observations in the census data and confirmed the model with mobile phone data. Similarly, Amini *et al.* compared the CDR data from different countries to infer the influence of social/cultural boundaries on human mobility [35, 42]. Despite criticisms regarding the potential sampling biases of CDRs [10-11], to date CDRs remain as one of the most comprehensive and versatile data sources in helping us understand large-scale human mobility.

Of specific interest in this domain is the study of human mobility in the context of our commute behaviors, as insights from such pursuits often have potent and far-reaching implications in urban planning, infrastructure construction, and even epidemiology. The first forays into this area using CDRs come from Becker *et al.* [12], who used the bulk mobile network data to understand the daily and nightly profiles of activities in Morristown, NJ. More recently, Issacman *et al.* [13] undertook a comparative study of daily commute patterns over two U.S. cities (New York and Los Angeles). While these studies have laid the groundwork for some key insights into the behavioral patterns in human commuting, they have been rather limited in scope. By focusing on a few select cities

Exploring universal patterns in human home-work commuting from mobile phone data specifically in the U.S., the similarities and differences observed are perhaps more accountable to regional determinants, as opposed to fundamental cultural and/or evolutionary factors. If we really wish to understand the characteristics of human commute patterns, we also need to set our eyes more broadly to countrywide datasets that come from different parts of the world. As a proof of concept, we have focused on Portugal (in Europe) and Ivory Coast (in Sub-Saharan Africa) for our study.

Of particular note, there exists a long-lasting debate about the universal uniformity of commute times, to which our analyses are poised to contribute. For example, in the prior studies on commuting patterns, Levinson and Kumar [14] noted that while commute speed and distance traveled may depend on the residential density, the effect on commute time seems ambiguous. Indeed, according to Kenworthy and Laube [15], and later to the 2010 American Community Survey [16], for American cities of various sizes (area and population), commute time seems surprisingly consistent at about 25-35 minutes. This is reflected in Figure S1, where we see that—despite wide differences in population size—mean commute times appear steady. A recent report by OCED in 2011 [17] echoes similar conclusions for various developed countries. Schwanen and Dijst [18], using the 1998 Dutch National Travel Survey, also proposed that the commute time as a fraction of total work time is roughly constant at 10%, which for an individual working 8 hours is about 30 minutes. Other studies have variously reported this "daily time constant" as 1.1 hours [19], 1.2 hours [20], or 1.3 hours [21]. This, in general, reflects what is known as Marchetti's constant, or what we will call the "constant travel time budget hypothesis", which posits that humans, since Neolithic times, budget approximately one hour per day on travel, independent of location, modes of transport, and other lifestyle considerations [22]. While we cannot easily infer individuals' commute behaviors that far back in time, an analysis [14] of datasets from Washington, DC from 1957-1988 suggests that the commute time is fairly stable at least within these three decades. Another camp of researchers has argued against this hypothesis. Goodwin [36], for example, reasoned that from the point of view of human psychology, a constant time budget would not be reasonable. Golob *et al.* [37] also pointed out that time expenditures tend to be inflexible in the short term (thereby giving rise to the apparent constant travel time observation), but more flexible in the long term. This view is corroborated by van Wee *et al.* [38], who showed that from various Dutch datasets, the commute time seems to have increased over the past decade. Similar conclusions were reached by Levinson and Wu, who studied commuting in the Twin Cities (USA) from 1990-2000 [39]. In addition, there are also studies showing that commute times not only vary by the cities within the same country [23], but also by the timing of the commute [24].

Mokhtarian and Chen [25], by agglomerating the findings from various studies, put together perhaps the most comprehensive review of the constant travel time budge hypothesis. They posited that travel time expenditures seem to change with factors such as income level, gender, and modes of transport. However, as the authors acknowledged, there still exists the possibility that travel time is constant over a city's entire commuting population (without subdividing the commuters into groups by mode of transport, income level, etc.). The researchers conceded that there may be significant limitations to their analyses, as their conclusions were borne out of a meta-analysis of diverse commute-related datasets, and as a result, there may very easily be confounding factors—such as survey types/questions [26], analysis units [27-28], types of trips included [29-30]—in how the data are collected/analyzed that would have influenced the observed outcomes.

We propose the use of mobile phone signaling data to minimize these possible confounding factors.. While there is no guarantee that individuals across all countries/cultures share the same call patterns, these cellphone datasets still contain some "common denominators" from which mobility behaviors have been inferred in previous studies. Therefore, given access to different mobile phone datasets at the country level (Ivory Coast, Portugal, and Saudi Arabia) and the city level (Boston), as well as a car-only GPS tracking dataset (Milan), we attempt to support/refute the constant travel time budget hypothesis. Investigating this hypothesis has important implications at the policy level, as it dictates how the population behaves when new modes of transport, roads, or other infrastructures are built [31].

Exploring universal patterns in human home-work commuting from mobile phone data

In this study, we focus on a specific type of commute known as home-work commuting. While the term "commute" may be more broadly defined to include any repeated trip between two or more locations, most of the studies cited above explore commute in the specific context of that between home and work. We first describe a methodology for inferring the home/work locations and aggregate commute patterns from mobile phone calls in different countries/cities, in comparison with the car-only GPS traces from Milan. While it is generally not possible for commute times to be accurately measured using mobile phone calls alone, the timing of the last call from home and the first call from the workplace is used (only for users who make frequent calls) as a proxy from which an individual's morning commute time can be gauged, and vice versa, for the evening commute time. We then test the methodology by investigating some interesting commute patterns. We close this study by testing a specific version of the constant travel time budget hypothesis with respect to people's commute behaviors. While this proxy for commute time, as defined above, generally results in an overestimation of the commute interval, we also describe approaches in which the actual commute time can be more accurately estimated in future studies.

**Materials and Methods**

We examined five different datasets. The first four sets ("Ivory Coast", "Portugal", "Saudi Arabia", and "Boston") are mobile phone signaling data, while the fourth set ("Milan") consists of GPS traces of cars. More details about these datasets can be found in the Results section below, as well as in Table 1.

*Spatial and temporal filtering*
For the cellphone-based datasets, because of the inhomogeneous arrivals of the calls from each user, in order to create some temporal uniformity required for Markov modeling, we employed time filtering by subsampling the data at 10-minute intervals. In instances where the inter-event period is more than 10 minutes apart, we assumed that the caller stayed at the original cell tower during this time period. Because for callers in areas with dense overlapping cell towers, it is a concern that the cellphone could randomly switch amongst different cell towers even if the caller actually did not move, we also employed spatial filtering using the same method as described in Calabrese *et al.* [32]. Essentially, we treated all movements within 1-km radius of the original cell tower as "noise" and counted movements that only exceeded this 1-km circle. In the case of the Milan GPS dataset, in order to make the data comparable to the CDR data, we first discretized the Milan Metropolitan Area into a mesh of grids 0.5 km by 0.5 km in size. The individual's GPS trace is accordingly discretized in the context of these grids, and applied a 1 km by 1 km low-pass step filter. If any user exhibited unusual mobility behaviors (such as moving at unrealistic speeds of 120 km/h or more), then typically the entire trace is discarded.

*Constructing individuals' travel portfolios*
Before we defined the home/work locations of each user, we first constructed an individual's "travel portfolio": a list of cell towers that the individual frequented, ranked by the time the individual spent in each place. Namely, the tower where the caller spent most of his/her time would be given a rank of 1, and so forth. This allowed us to construct a profile of each user's movement as he/she moved from a tower of a certain rank to a tower of a different rank, or spent a certain time in a tower of a given rank. In the case of the Milan GPS dataset, having obtained each user's GPS trace in a series of grids, we then identified the most frequently visited grids (measured by the cumulative time spent in each grid). Then the frequently visited places were ranked at the grid-level resolution. The sample GPS trace of one user is shown in Fig. S2(b) as an illustration of this method.

*Identifying home/work locations*
To identify the home/work locations, we first filtered out calls on Saturdays and Sundays, because in the countries/cities of study, people tend not to go to work on these days of the week. Due to the fact that weekends are defined differently in Saudi Arabia, we filtered out all activity on Thursdays and Fridays in the STC dataset. Then we filtered out the call sequences that are too infrequent to give a meaningful estimate. To do this, we only considered sequential calls that are spaced less than 16 h apart, and assumed that after a call, the caller stays at the same location until the time of the next call. From this analysis, we were able to assign the total daytime and nighttime periods spent in each location, with the thresholds set at 8 a.m. and 8 p.m., respectively. If an inter-call period happened to span across the day/night thresholds (of 8 a.m. and 8 p.m.), then we split the interval at the threshold and assign the daytime and nighttime intervals correspondingly. For each user, we identified the daytime and nighttime locations in which the user spends the maximum dwell-time, as long as such location accounts for more than 50% of the total observed daytime and nighttime dwell-times in the user's travel portfolio. If such pair of day/night places existed, then we assigned these places accordingly as the home (night) and work (day) locations. If such places do not exist for the user, we disregarded the user. Our method is roughly similar to that used by Phithakkitnukoon *et al.* [33], but we imposed a more stringent filtering by requiring each user to spend more than 50% of the total observed daytime/nighttime dwell-times for the place to be identified as the work/home location. While this stringent filtering ensures that spurious signals are minimized, it also limits the scope of our study by excluding certain occupations without a fixed location at a fixed time of the day (such as salesmen, drivers, etc.). While it is certainly possible that people with these occupations still have well-defined office spaces and homes, their home/work locations will be much more challenging to identify using our methodology described above. Insofar that these people consist of a small fraction of all commuting population, we have chosen to ignore them from our study. In brief, our analysis is for steady working people with clear home and work locations. In general, due to our stringent filtering method, about 7% (Portugal) to 11% (Ivory Coast and Saudi Arabia) of all users available in a dataset will have well-defined home/work location pairs from which further analysis can proceed.

In the Milan GPS dataset, the home/work locations are estimated in a similar manner, with the assumption that the individual always stays in proximity to his/her car. This assumption is not always valid, as the individual may park his/her car and run several errands at the same time. Despite potential inaccuracies that arise due to these behaviors, the nature of the GPS dataset makes accounting for such behaviors impossible, unless the same dataset can be overlaid user-by-user against other mobility-related datasets (such as a CDR dataset of Milan, which was not available to us).

*Computing commute distance*
Once the home/work locations are identified, then we computed the commute distance as the great circle distance between the home and work locations. There are many approaches in literature to estimate the commute distance: for example, the crow-fly distance (either the great circle distance or the Euclidean distance) [44], the shortest distance path (SDP), or the shortest time path (STP) [45]. As CDR datasets are unable to exactly reproduce the routes of commute, unless coupled with GPS traces or further questionnaire information, we chose to calculate the crow-fly distance. Of the two approaches (great circle versus Euclidean distances), the former seems to be more accurate especially in cases where the commute distance is long. We realized that the great circle distance is not the most accurate measure of the actual commute distance. Depending on the modes of commute (e.g. bus, train, car, etc.), the correction factor between the actual commute distance and the great circle distance may differ. In general, for a commute distance greater than 5 km, this correction factor is quite consistent at about 1.3-1.4 for different modes of transportation [46]. However, below 5 km, this factor can either increase drastically (for cars), or decrease (for public transport). Since the CDR datasets likely include mixed modes of commute, we can at best say that the under-estimation for medium/long commutes (~ 5 km or more) will be consistent, which will impose a systematic correction factor on our great-circle commute distance. For shorter commutes (i.e. < 2.5 km), this error may become significant and dependent on modes of transport, and this may limit our ability to take accurate commute distance measurements at short distances. However, with the lack of data on other details about individuals' commute, here, we simply take the great circle distance as a proxy for the commute distance, which will be always greater than the great circle distance.

As a validation of our method, when we calculated the commute distances in Portugal, our results are qualitatively consistent with those found in Phithakkitnukoon *et al.* [33] from a different Portugal dataset, even after we applied more stringent filters as described above. As further verification of the validity of the home-work distance definition, we also tested for correlation between an individuals' commute distance and his/her radius of gyration (calculated using the method described in González *et al.* [7], and we found statistically significant correlations between the two quantities. Because most of the datasets span a relatively short period of time, we did not account for the possibility that the phone user moved his/her home or work location during the period, and assumed that such events are rare.

*Determining morning/evening commuting times*
To estimate the commuting times, we adopt the follow algorithm. For the morning commute times, for each user, we identify the timestamps of the last call from the "home" location and the first call from the "work" location before noon. The difference between these two timestamps gives us an upper estimate of the morning commute time. Likewise, for the evening commute times, for each user, we identify the timestamps of the earliest call from the "home" location and the last call from the "work" location after noon. Once again, the difference gives us an upper estimate of the evening commute time. To ensure that our calculated commute times do not trivially reflect the calling frequency, we further filtered the dataset so that only those who called sufficiently frequently (on the order of one call per hour on average) were included in the final reckoning. These stringent did filter out more than 95% of the usable dataset, though even so, we were still able to locate about 20000 (Ivory Coast), 50000 (Portugal), to 260000 (Saudi Arabia) users with identifiable morning/evening commute times. Furthermore, we also concede that in this case, our calculated commute times may still be distorted by the fact that people may not immediate make calls right before leaving or arriving home/work, and those who make infrequent calls may appear to have a somewhat larger calculated commute time. However, given that the previous literature (see the introduction section above) that commute times typically fall within 30-60 minutes in duration, and given that we only consider users whose frequency of call on average are on the same order of this duration, then if we do indeed observe moves between home/work locations in the same timeframe, we do not expect latter factor (inter-call times) to dominate our estimation.

Despite these filtering and correction measures, there are still some potential limitations about the datasets that can negatively affect the accuracy of our results and their interpretation. Such included the differences in the mobile phone usage behaviors across different users, and notably, across different countries/cultures. People who call at different times of the day with different frequencies, for example, can affect our estimations of the commute times in different locations. There exist more elaborate correction mechanisms that we may employ to further screen for these confounding factors. However, given the limited size of the dataset already (for example, for Ivory Coast, about 500,000 total users, each traced over only 2 weeks), such more elaborate methodologies are beyond the scope of our study. However, in the Discussion section, we review in greater detail the potential impacts that these limitations may have on the accuracy of our study of human commute behaviors.

**Results**

Having developed a common way to parse for home/work and commute information that can be equally applied to different datasets, we can then ask what insights this methodology can reveal to us regarding human mobility and commuting. In this section, we discuss a few topics/insights about the datasets that stem from our methodology, and conclude by focusing on testing the constant travel time budget hypothesis in the context of commuting.

*Data description*
The first dataset ("Ivory Coast"), provided by Orange telecom, spans 150 days from December 1, 2011 to April 28, 2012, and consists of the consecutive call activities of 50,000 randomized subscribers and is provided as a part of the Data for Development (D4D) Challenge. Each record in this dataset has the following output: timestamp, de-identified ID of the user, and the antenna of connection (one of the approximately 1,200 antennas in Ivory Coast). This data set is broken into 10 subsets, each of which track 50,000 different subscribers over a two-week period. The 10 subsets are consecutively ordered by the time period so that taken together, they span the entire period of study, though the 50,000 subscribers in each dataset are re-randomized to ensure anonymity. The second dataset ("Portugal"), also provided by Orange telecom, spans 2 years from January 1, 2006 to December 31, 2007, and is similar in its output as the Ivory Coast data set in the context of Portugal. The size of this set is 400 million CDRs from 2 million users, and has about 6,500 antennas. The third dataset ("Saudi Arabia"), provided by Saudi Telecom Company (STC), covers the entire country of Saudi Arabia, with approximately 100 million daily network connections to over 10,000 unique cell towers. The total dataset encompasses roughly 14 million devices. Each individual record holds the caller's location, precise time and duration measure, type of connection, and type of service (subscription, pre-paid, etc.). The fourth dataset ("Boston"), provided by AirSage, consists of mobile device signaling data in the Greater Boston area from 2 million mobile devices over the course of 4 months from July to October 2009, containing about 900 million records per month. Finally, the fifth dataset ("Milan"), provided by Octo Telematics, consists of GPS traces of cars (about 99,000 cars and 18 million positions) in the Milan metropolitan area over a period of one week. Sample GPS traces and density plots are shown in Fig. S2. A summary of these datasets is given in Table 1. All data submitted have been anonymized prior to receipt and are in line with all local data protection laws.

*Location resolution*

There are some concerns of whether or not the distribution of cell towers in our datasets offer sufficient spatial resolution to interrogate human commuting behaviors, especially for countrywide datasets such as Ivory Coast, Portugal, and Saudi Arabia. Figure 1 in a previous study by Amini *et al*. [42] attempted to characterize this spatial distribution in detail. Figure S2(c) in this paper plots the cell tower density for Saudi Arabia. Notably, the cell tower spacing is not uniform throughout the countries; rather, as expected, they are most concentrated within urban areas, often with an inter-tower spacing of less than 1 km. In rural areas with very sparse population density, the spacing amongst cell towers can be more than 100 km apart. This spatial inhomogeneity may pose concerns regarding the spatial accuracy of our home/work commuting characterization, especially in rural areas with large inter-tower spacing. While in CDR datasets, unlike GPS traces for example, typically the spatial resolution is beyond our control, we argue that commute is most interesting and relevant in urban and semi-urban areas, where cell towers in both Portugal and Ivory Coast are quite uniform and closely packed. For the rural commuters (for example, into/out of small provincial towns), the sparse spacing in rural areas may cause inaccuracies in two accounts: (1) it can grossly over-estimate the commute distances of people whose commute happens to cross from one cell tower to the next with a large inter-tower distance; and (2) it can grossly under-estimate the commute distances of people whose commute does not cross cell towers (whose home/work location would be identical). If it were possible to assume that everyone commutes, then on the aggregate population level, this error could be averaged out, giving rise to an estimate that approaches the true population mean. However, without this assumption, it became harder to accurate estimate rural commutes, without the aid of further datasets such as GPS traces on smartphones. Given the limited datasets available at our disposal, we did not undertake a detailed quantification of this effect, except noting that the people who undertake rural commuting make up a minority (in both countries, less than 5%) of the overall countrywide commuters. In our study, we filtered out users who spend significant time in a cell tower that is more than 50 km from adjacent ones, though in a future larger study with complementary datasets (such as smartphone GPS traces), there will be more elaborate measures that can be taken to ensure accuracy of this minority group of commuters.

On the other hand, while, as discussed above, GPS coordinates are generally more spatially accurate for quantifying human mobility compared to cell tower locations, it is still conceivable that in some areas (such as tunnels or under buildings), the ability to detect GPS signals may be impaired. If so, such locations may be under-represented in our data. However, we argue that these circumstances, in daily commuting conditions, are typically rare. As shown in Fig. S2(a), for reported GPS coordinates aggregated over one day in Milan, we can generally see clear delineation of roads, which is what we expect.

*Individuals display limited travel range during the night*

Exploring universal patterns in human home-work commuting from mobile phone data

In the process of computing an individual's home/work locations, it is first necessary for us to determine the individual's set of frequented locations. So before we analyze home/work commuting, we first make an interesting observation on the effect of day/night on people's range of travel. Here, "range" is defined specifically as the size of the set of non-overlapping locations visited by an individual with a non-zero frequency over the observed period. Using the analyses described in the methods section, we quantified how much time (as reflected by the Markovian self-transition probability) people spend in each frequented cell tower, and rearranged the cell towers by their rank number (corresponding to the total dwell time). Figure 1 below shows the average dwell times in locations of different ranks in an average person's travel portfolio in Portugal (left, red) and Ivory Coast (right, blue), categorized by daytime activities (solid lines with closed dots) and nighttime activities (dashed lines with open circles), and plotted on a log-log scale. First, from the daytime activities, we notice that the distribution of dwell times roughly follow Zipf's law, with comparable power law coefficients. This is consistent with prior observations, such as in González *et al.* [7]. However, remarkably, during the night time, the distribution of dwell times show a distinct change in both countries: instead of following Zipf's law, the distributions assume a sigmoidal shape, with a sharp fall-off at around a rank number of 10 (as shown in the plots below where the two curves intersect). This suggests that while during the day, people are active over a wide range of places (represented by different cell towers), during the night time, people tend to limit their travel destinations by visiting mostly the major ones. This behavior is consistent in both Portugal and Ivory Coast. Other datasets show similar results.

*Different locations show distinct commuting distance profiles*
Next, we focused on the individual home and work locations and aggregate commute behaviors in Ivory Coast and Portugal. We characterized the home/work locations for the different users based on the technique outlined above. We then estimated each individual's commuting distance as the distance between the home/work cell towers, and plot the probability density function below in Figure 2, with the inset showing the tail-end behaviors at long distances plotted on a log-log scale for the different datasets in different colors.

To better make sense of Figure 2, we first focus our attention on Ivory Coast, Portugal, and Boston, and will come back to the special case of Milan. As can be seen in Figure 2, the distribution of commute distances are significantly different for commute distances of less than 10 km: there are significantly many more people in Ivory Coast who live very close to their work place compared to Portugal or Boston. And yet, for longer commute distances (more than 10 km), all datasets exhibit similar Zipf's law behaviors, with the exception of the Boston dataset. However, when we performed a two-sided Kolmogorov-Smirnov K-test on the null hypothesis that the commute distances between any pair of dataset are drawn from the same distribution, the null hypothesis was rejected at a very significant level ($P < 10^{-11}$), demonstrating that the distributions, overall, are quite different in nature. This is also qualitatively observed in Figure S4, which shows significantly different cumulative distribution functions across the different datasets. As a crude quantification, we calculated the mean commute distances of the two countries, and found Ivory Coast to be 20.2 km, while Portugal to be 25.4 km.

This slight difference, coupled with the earlier observation that a much larger fraction of Ivorian's live close to their work places, is suggestive of the existence of two distinct commuting populations. The first group, which we will call "long commuters", consists of people who can afford to live far away from their work places (> 10 km). In Ivory Coast, due to the limited public transportation infrastructure, members of such group likely possess their own means of transport (e.g. a car). The long commuters' behavior is very similar between Ivory Coast and Portugal. On the other hand, the second group, which we will call "short commuters", consists of people who live closer to their work places (< 10 km). Because of the complications of owning their own cars in an urban environment (e.g. parking, traffic jams), we surmise that this group is more likely than the "long commuters" to rely on the public transport or just commute on foot. While in the Portugal and Boston the public transport is rather well developed and wide-reaching in these regions, which enables people to live further away from the city and still be able to commute in a timely fashion. While this explanation is

Exploring universal patterns in human home-work commuting from mobile phone data speculative without further supporting data, it is nonetheless consistent with the observations above in Figure 2.

In the case of Saudi Arabia, we see that the distribution of commute distances again diverges significantly from Portugal and Boston under about 4 km. Similarly to Ivory Coast, many more individuals live closer to their places of work.

Finally, Milan represents a different case in which the dataset is GPS traces from cars instead of phone signaling data and therefor represents only a subsample of all the commuters - drivers. Here, while the initial distribution is qualitatively similar, the long tail falls off at a different slope, as shown in the inset. This is likely the effect of sub-selecting the mobility pattern in which individuals commute by cars. We see that in the long distance regime of above 30 km in commute distance, such commute distances are less frequent than the other aggregate mobility datasets. This may simply reflect the fact that it becomes less economical (from the perspective of time, fuel, and labor) to operate a car over long-distance commutes, in preference for other modes of transport (such as commuter trains) that may be available in the local context.

*Portugal and Ivory Coast show distinct commute timing characteristics*
Beyond the bulk commute distances, we are also interested in further examining any distance-dependent behaviors such as the timing and time interval of commuting in both countries. Using the method described in the Materials and Methods section above, we first computed the timing of the morning/evening commute for the two "standard" countries in our datasets, namely Portugal and Ivory Coast. Because CDR data cannot accurately tell us when exactly a person is making a trip, we used the last call from the home cell tower in the morning as a proxy for the timing of the morning commute. Similarly, we used the first call from the home cell tower in the evening as a proxy for the timing of the evening commute. We then binned the individuals by their commute distances (< 2.5 km, 2.5-5 km, 5-10 km, 10-20 km, and 20-50 km), and plot the distribution of the timing of morning/evening commutes in Figure 3 below.

As seen in both plots, we are able to capture the commute peaks in the morning (around 8-10 a.m.) and in the evening (around 8 p.m.) in both countries. The peak patterns above are in agreement with the characteristics observed in the traffic congestion model proposed by Vickrey [34]. We noted that these distributions can be reasonably fitted to Gaussian distributions, if we make the assumption that for a population at a given distance from his/her work place, the time that each individual makes his/her commute can be approximated by a random variable with a normal distribution. Fig. S3(a) outlines this Gaussian fitting approach in greater detail, and in Fig. S3(b), the goodness of the Gaussian fit is quantified as a Q-Q plot. As we can see, for the typical commute domain (for example, around 5 a.m. to 10 a.m. in the morning), the Gaussian distribution is a reasonable fit. Outside of this domain, there are generally minor deviations: the CDR-derived distributions show a "shorter tail" compared to a theoretical Gaussian distribution (which extends to infinity). This makes intuitive sense, because in general, people commute in a limited time window (for example, it would be extremely rare to find people commuting from home to work at 10 p.m. at night). Furthermore, our definition of home/work location by the most frequented night-like/day-like locations also excludes the possibility of commuting at arbitrary times of the day beyond a certain reasonable morning/evening window.

In order to better quantify the inherent differences in the timing of commute in different regions, we measured the peak commute times in each of the distributions above as a function of the commute distances. In order to calculate the peak commute times, we undertook two different methods. In the first method, we equated the peak commute time for each distribution to the median time from the entire distribution. This method would minimize the influence from extreme outliers (such as the low-level of activities as early in the morning). In the second method, we first fitted each distribution to a Gaussian distribution, and then equated the peak commute time with the mean of the fitted Gaussian. For each method, we plot results below in Figure 4 for morning commute (left column) and evening commute (right column), calculating the peak times using the median time method (first row) and the fitted Gaussian method (second row).

As observed in Figure 4 above, first we note that regardless of the method of calculation, the existing trends are quite reproducible. In particular, there is quite a significant dependence of when people leave home in the morning as a function of the commute distance in both countries. As expected, the further people live from their work places, the earlier they opt to leave home in the morning. If the constant travel time hypothesis were true, then, this would also imply that people who live further from their work places would also arrive at work earlier. We then attempted to quantify the significance of correlation through the Spearman's rank correlation test, which is chosen because the *x*-axis is qualitative (showing a range of commute distances) while the *y*-axis is quantitative. As summarized in the test statistics in Table 2, the correlation tests for the morning commute timing (based on either calculation method) as a function of commute distances show significant negative correlations for both Ivory Coast ($\rho$ = -0.92) and Portugal ($\rho$ = -0.93) at the 2% significance level.

One caveat for performing the statistical correlation tests here is that given the limited sample points (5-6 data points per plot), the results from such tests are generally not useful. We therefore do not seek to draw strong conclusions from any potential correlations (or the lack thereof) above regarding the commute times, but rather only describe these relationships phenomenologically, given the limited datasets. In the future, if the availability of larger datasets over longer periods can yield more finely resolved commute timing as a function of commute distance, then the same approach described here can be applied with greater efficacy and statistical significance to evaluate any observed statistical correlations.

On the other hand, in the evening, the situation is much less clear-cut. In Portugal, there is a weak position relationship between how late a person arrives home as a function of how far he/she lives from work. In contrast, in Ivory Coast, the statistical tests are unable to show a significant relationship at the 5% significance level. While, as explained above, due to the small sample size we cannot draw definitive conclusions, if we assume that this lack of significant relationship is indeed true for Ivory Coast, then it may suggest that regardless of the commute distances, Ivoirians tend to arrive home uniformly between 8-8:30 p.m. in the evening. If this is true, then one potential explanation for this difference in commute behaviors between the two countries is the differences in commute conditions. It can be argued that in Ivory Coast, the commute options are much more limited, because (1) the limited availability of public transport, and (2) the hazardous night driving conditions. Therefore, after nightfall, people are compelled to reach home by a given time, regardless of how far they live from work. In contrast, in Portugal, the availability of public transport and adequate road lighting and safety conditions mean that people may feel more comfortable staying later. This may account for the absence of clear dependence on the time of arrival at home as a function of the commute distance in Ivory Coast: regardless of how far/close a person lives, there is a pressure to get home by a certain time. Without the support of further transportation data specifically from Ivory Coast (in comparison with Portugal), we acknowledge that the explanations above remain mostly speculative. However, in future studies, it will be interesting to couple CDR datasets with other data on public transport and road conditions in order to better understand and quantify the underlying mechanism that drives observed differences in commute behaviors.

*Commute time appears invariant with commute distance*
Having examined the timing characteristics of commuting in Ivory Coast and Portugal, we then proceeded to parse out the commute time interval: the time it takes for an individual to get from home to work in the morning, and vice versa in the evening. Once again, due to the challenges in inferring the exact departure/arrival times from home/work from CDRs, we utilized the proxy as described in above in the Materials and Methods section. We then binned the users once again by commute distance in each dataset/region. We also added the Saudi Arabia dataset as a countrywide comparison, Boston as a city-level comparison, and Milan as a comparison using a different type of dataset (GPS traces).

Figure 5 illustrates the mean commute times of the users binned by increasing commute distances. We first focus on the mobile phone datasets (Ivory Coast, Portugal, Saudi Arabia, and Boston), and will discuss about Milan later. We first observe that the characteristic commute time can vary from place to place and from morning to evening, for the call-record-based datasets in Ivory Coast, Portugal, Saudi Arabia, and Boston, which seems to support the view that a universal "Marchetti's constant" does not really exist. Nonetheless—with the exception of possibly long commuters in Saudi Arabia—we see that the mean commute time, remarkably, does not change significantly as a function of commute distance, as most fluctuations lie within the error bars (with important caveats about short commutes to be described later). This seems to point to a location-dependent invariance in commute time, even though at a more universal scale (across different countries), an invariant commute time may not exist. The main anomaly in Figure 5 appears to be the GPS-based Milan dataset (on cars only), which shows a monotonic increase of mean commute time as a function of commute distance. This seems to make intuitive sense as and appears consistent with Mokhtarian and Chen's observations earlier [25]. In essence, if we only examine cars (as is the case of Milan's car GPS traces), the naturally we expect that the further people have to commute, the longer the driving time will be. In contrast, the CDR datasets are agnostic about the method of commuting: and in that sense the different types of commute behaviors are aggregated. Namely, if a person lives close to the workplace, he/she may choose to walk or bike rather than to drive. If a person lives close to a commuter train station, he/she may afford to live further away from the workplace without suffering the consequences of long daily drives by car. This is in fact one formulation of the constant travel time hypothesis: depending on the means of travel available, people tend to adopt lifestyles where the time they spend on travel (i.e. commute) is approximately constant. While this explanation still remains a hypothesis, it is further supported by data from Saudi Arabia. It is important to note that Saudi Arabia, a country known for poor public transportation infrastructure, encourages personal vehicle use through highly subsidized petrol prices [47]. In Riyadh, the country's largest and arguably most developed city, only 2% of daily trips rely on public transportation [48], a figure that stands in sharp relief to Lisbon's 28% ridership [49]. One could argue that—due to the country's comparatively high reliance on car transport—the dataset exhibits a slightly Milan-like (car only) behavior where the mean commute time appears to have some positive dependency on the commute distance, even though in this case, the CDR records and not the GPS traces are being analyzed. Desiring to characterize these observations better, we proceeded to plot the distribution of commute times associated with each distance bin in Figure 6.

Figure 6 shows the plotted outcomes in the three call-record-based datasets, for the morning commute times, for the subpopulations in each country with different commute distances. The first key feature to note is that different countries/regions exhibit different shapes of distributions. Ivory Coast (a) and Portugal (b), for example, have a particular commute time distribution that has a peak around 30 minutes. On the other hand, Boston (c) has another characteristic shape that falls off sharply. Such differences in distribution shapes imply that the main characteristics of commute time may differ from place to place. This is supportive of our earlier observation in Figure 5 that different places have different mean (characteristic) commute times.

The second key feature to note is that, beyond minor differences, the distributions otherwise show remarkable similarity that is independent of the commute distances, especially for Portugal, Ivory Coast, and Boston. This implies that there exists some overall distance-independent law that governs the distribution of commuting behaviors in different local contexts (countries/cities), despite that such a law seems to vary from locality to locality (as observed in the different characteristic shapes of distribution in different regions). In layman's terms, an individual in Boston and an individual in Ivory Coast may have different concepts of "acceptable commute time". However, within a given region, different individuals—despite differences in income levels, available methods of transportation, etc. will adapt their lifestyles such that everyday the time they spend traveling is within the "acceptable commute time" range in their specific region/context. In other words, even though a universal Marchetti's constant does not seem to exist across different locations, at each local country/city level, such a constant may well exist within a limited jurisdiction.

Finally, we also made an attempt to compare call-record-based data (which should encompass mobility in general) with car-GPS-based data (which should only sub-select mobility specific to car transport) by plotting the same morning/evening commute time distributions for the Milan dataset (d, i). In this case, interestingly, the earlier commute distance independence no longer holds true, as the subpopulations having different commute distances also have different distribution shapes of commute times. As is consistent with observations made in Figure 5, this does make sense, because if we choose to focus only on car drivers, then it is clear we expect that people who live further away may have to spend more time driving. In contrast, if we focus on the aggregate data (represented by call-record-based data), then this difference is mitigated by the fact that people living in different distances from work have the option of selecting different modes of transport (such as cars, trains, bicycles, etc.), so that their commute time is minimized. If we examine Saudi Arabia's commute time distributions, we once again see that it falls somewhere between the distance-independent case of Portugal, Ivory Coast, and Boston, and the distance-dependent case of car-only Milan. This in part also vindicates Mokhtarian and Chen's earlier position[25], that the constant travel time budget hypothesis applies only at an aggregate level, when we can give people enough choice in the mode of mobility, and given so, people figure out ways to minimize their commute time/effort, and overall, this minimized time seems to be consistent in different contexts. However, what is new insight in our study is that beyond a certain level of aggregation at a city/country locality, a search for an even more universal commute behavior (across different countries/continents) seems to suggest that it does not to exist, at least on the basis of CDR datasets and our specific methodology.

To analyze this pattern further, we note that the evening commute time distributions (plotted in the second column of Figure 6) show a slightly more uniform shape compared to the morning distributions, suggesting that in the evening, people take longer to get from work to home. This makes sense, as in the morning, people often head directly to work from home, whereas in the evening, there are more intervening opportunities in terms of dining out, running errands, shopping, etc. However, even in the evening commute scenario, the distributions seem independent of commute distance across the same "category" of datasets (full-fledged countries versus cities), which is consistent with the observations above, except for the special case of Milan (where GPS car-tracking data, rather than bulk mobile-phone-based commute-based data, are used), and possibly Saudi Arabia (heavy reliance on cars).

It is also acknowledged here that Figures 5 and 6 contain some observations which may appear, at a first glance, to be unphysical and to raise doubts about what exactly CDR data can actually measure about commute time intervals. For example, people with a commute distance of less than 5 km still has an average commute time more than one hour in some cases, which seems unrealistic. We acknowledge this issue and will address it fully in the Discussion section below.

**Discussion**

In this study, we started with the premise that the difficulties with verifying/disproving the constant travel time budget hypothesis (as outlined in the introduction) often lie in the comparability between the different datasets, where confounding factors such as data collection/analysis may exist. We proposed to minimize this effect by finding a common approach of quantifying commuting through mobile phone datasets—from which reasonably comparable commute characteristics can be inferred from different countries/cities of interest. We examined four call-record-based datasets (Ivory Coast, Portugal, Boston, and Saudi Arabia), and also compared these with one GPS-tracking-based dataset (Milan). We described a methodology for inferring the home and work locations for different users, as well as for computing the distance and the timing of commute. While certain assumptions remain in our methodology (for example, by arbitrarily assigning the day/night boundary at 8 a.m./p.m. for defining home/work locations), we argue that because these assumptions were applied uniformly across all datasets, any systematic bias should not affect the comparison of the outcome.

Exploring universal patterns in human home-work commuting from mobile phone data

As a proof of concept, we computed the commuting distance as well as the commuting timing from these datasets, which are in agreement with known characteristics/models from existing commute studies. Despite the differences in these variables across different datasets/locations, when we plotted the distribution of commute times, we found remarkable distance-independence across the call-record-based datasets of Ivory Coast, Portugal, and Boston.

This stands in contrast with the car-only GPS-tracking-based dataset of Milan and the car-heavy CDR dataset from Saudi Arabia. The Milan dataset shows strong dependence of commute time as a function of commute distance (as those who have to drive further typically commute for a longer time). The Saudi Arabia dataset is seen as a mixture of the two extremes, as while the CDR nature of the dataset may help aggregate the different modes of commute (e.g. walking, bicycling, public transportation, private cars, etc.), we also know that the country is heavily car-dependent as discussed previously. This implies that the constant travel time budget hypothesis (as it pertains specifically to morning/evening commuting) holds true only at the aggregate country/city-wide level of mobility (where there are choices of different methods of commute available to the individual depending on the distance, location, etc.), which validates Mokhtarian and Chen's hypothesis [25]. However, it is also true that our analysis did not conclusively identify a universal "Marchetti's constant": despite their localized independence on distance traveled, characteristic commute times seem to vary due to location (Portugal, Ivory Coast, Boston, and Saudi Arabia), as well as due to time of the day (morning and evening). Therefore, in light of our findings across the five datasets, we can propose here as a testable hypothesis what we call the "localized form" of Marchetti's constant: even though in different regions, people may have different commute time characteristics (dependent, for example, on the cultural perception of time, the overall infrastructural development of the country, etc.), if we focus on a single region, then we find that most people display commute time characteristics (e.g. average time, distribution of times) which tend to be independent of the commute distance. As an illustration, individuals who can afford to drive may live in places not within reach of those who walk, bicycle, or take the public transport. Individuals who live close to commuter rail stations may live in places that may have too many traffic jams to be acceptable for car commuters. In other words, individuals may distribute themselves geographically and adopt their lifestyles (e.g. commute behaviors) in a way so that they spend reasonable amount of time of their lives commuting.

Ultimately, CDR datasets, like other datasets, are not perfect. As demonstrated above, our methodology is able to reveal existing and salient patterns in commuting behaviors in different regions/contexts. However, at the same time, we should discuss the caveats and limitations of the CDR datasets. As we raised earlier in the Results section, the reader may have questions about the accuracy of CDR data in estimating commuting times, as Figures 5-6 produced, on the basis of our CDR analysis, some unphysically long commute times even for short commute distances of 5 km. To properly address this question, there are three inter-linked questions that need to be discussed in sequence below. The first question is: What does an actual home/work commute constitute? While it is true that some people will travel directly from home to work, and vice versa, the likelihood is that many people will also make intermediate stops. Yang *et al.* [40], by studying the commute data from Shangyu, China, demonstrated that these intermediate stops—which occur frequently at commercial establishments (for example, restaurants for breakfast/dinner), can be a significant part of the commute. Likewise, Schneider *et al.* [41] by quantifying the motifs in daily trips using CDR data, also discovered various means by which individuals incorporated these stops/detours. If it is true that such stops are common, then irrespective of the toolsets/datasets used to interrogate commuting, they will confound the accounting of commute times because (a) of the extra time spent at intermediate stops, and (b) of the extra detour taken from the most direct home-work path in order to reach these stops. While it may be possible to quantify these detours using adequately large CDR datasets, it is still a rather challenging endeavor, given that the mobile calls may be too sparse to identify stops consistently. Even if a caller makes frequent calls between home and work, given that a mobile phone call at an intervening cell tower between an individual's home and work locations may indicate one of two things: (a) the individual has made a brief intermediate stop, and (b) the individual is simply calling en route without stopping (unless this particular intermediate cell tower is obviously and consistently out of the way between home and work locations). Given our methodology and the current scope of our study, we did not explore this topic further, but rather lumped everything into the overall commute time. This is one reason why some commute times may appear unreasonably overestimated on the basis of CDR records.

The second question is: What exactly about commuting time can CDR data measure? As mentioned earlier, the dataset does not guarantee that an individual will always call immediately before he/she leaves from home/work, and immediate after he/she arrives at home/work, even though often, people may call before leaving for a trip so others know that they are on their way, for example. All we know for certain is that in the morning, after the individual has left home for work, then all calls from the home location should cease. Likewise, there should be no calls in the morning from the work location before the individual has actually arrived at work. Therefore, what we measure as a commute time on the basis of CDR data is simply a proxy of the actual commute time; it will certainly overestimate the actual commute time based on how frequently an individual calls. The best we did to address this in our data processing step is to only select individuals who call frequently enough, at least during the particular period of morning/evening commute. However, even so, there are variations in the frequency of calls from individual to individual. For example, some people may prefer to make calls while in transport, while others do not. Such variations in calling habits may also vary from country to country depending on the cultural context. If we had access to larger datasets over many regions tracing over individual call patterns over a long period of time, then there are tangible ways to control for these variations in our commute time estimation. However, in reality, we only had data available for a limited number of individuals over a limited period (for example, in the case of Ivory Coast, 2 weeks), and this has made further quantifications/controls challenging. In the future, as more datasets become available over a longer period of time, it will be worthwhile to revisit these questions in order to improve the accuracy of commute time estimations.

Furthermore, the third question is: how accurately can our methodology detect short commutes? One limitation on our methodology is that the application of a spatial filter of 1 km, which while limits the noise due to random switching of cell towers especially at the boundaries between two towers, also causes our study to ignore those whose commute distance is less than 1 km. More importantly, as discussed above already in the Methodology section, what we can measure from the CDR data is not the actual commute distance, but rather the great-circle distance between home and work (as the crow flies). As mentioned above, while for medium/long commutes, the difference between these two quantities can often be consistently and systematically corrected with a factor, this is not the case for shorter commutes, where the correction factor may depend on the types of transportation, for example. Depending on the magnitude of the correction factor for short commutes, this may have caused us to either over-estimate or under-estimate the commuting times of short commuters with respect to the commute times of long commuters. Supposing that this were true, then it would raise doubt about the constant travel time hypothesis at short distances (i.e. within walking distance). However, there exist other equally salient explanations for these hour-long short commutes that may nevertheless still be consistent with the constant time budget hypothesis, as mentioned earlier: intermediate stops on the way to work, heavy traffic conditions in the city (if we assume that short commuters live close to their work places, which are more likely to be located in the city than the countryside), as well as the fact that individuals may not immediate call before leaving home or after arriving at work. In order to get a more accurate resolution for short commuters, it will be necessary to couple the CDR dataset with other techniques with a greater spatial resolution, such as GPS traces on mobile phones, so that we obtain the precise routing of the commute (to calculate actual commute distances) and to more accurately gauge the home-departure and work-arrival times.

Given these limitations, does CDR fare better in relation to other mixed datasets used earlier in interrogating human commute behaviors? Many of these mixed datasets, as mentioned earlier, are surveys of individuals in different cities. One generally can expect that these self-reported commute times give quite an accurate measure, if controlled properly and reported in an unbiased manner. However, as already described in previous meta-analyses [25], most of these survey-based studies are only available in limited locales; any attempts to study commute behaviors across different regions/countries are currently often confounded by factors such as variations in survey design and

Exploring universal patterns in human home-work commuting from mobile phone data implementation. If we were to carry out a global-scale study of human commute, it would also be unrealistically costly to obtain reliable data in different countries for a comparative study. We see CDR datasets—which come from mobile phone users throughout the world—as having the potential to overcome such limitations. While, as discussed above, the nascent and limited availability of CDR datasets may have raised questions about the accuracy of our commute characterization, especially at short distances, we believe that in the future, with the increasing accessibility of larger datasets, there are methods and correction measures that can be implemented to minimize such inaccuracies.

Ultimately, have we provided an answer for the age-old debate about the constant travel budget hypothesis, or the Marchetti's constant? From our analysis above, we can say that in each location that we analyzed, there seems to be remarkably distance-independent commute behaviors for medium/long commutes (> 5 km). We cannot reliably conclude for shorter commutes due to limitations on the CDR data.

However, when the commute times are averaged across the whole, we did not identify a characteristic commute time that is invariant across all locations. While this may be an argument against the existence of Marchetti's constant, it may also be that given the limited datasets, we did not have the wherewithal to properly account and correct for the different mobile phone usage behaviors across different locations, which could also have impacted our estimation of characteristic commute times in these different locations. Hence, while we did not conclusively answer the question regarding Marchetti's constant, we believe that with increasing availability of larger CDR datasets in the future, this question will be highly interesting and relevant to revisit, perhaps with the implementation of better correction measures to account for any intrinsic differences in mobile phone usage behaviors. We also recommend that future studies couple the CDR datasets with other supporting non-CDR sources.

Overall, in this paper, we developed a methodology that allows us to interrogate human commute behaviors, and applied it to several mobile phone and GPS datasets. The fact that we were able to observe common commute features despite the highly diverse nature of these datasets offer a compelling demonstration that there are some aspects of human commuting that are universal, and we consider this to be a novel development of the use of mobile phone datasets to better understand commute behaviors. Comparisons between car-only (Milan) or car-heavy (Saudi Arabia) datasets to other datasets which are more agnostic to methods of commute (Boston, Portugal, and Ivory Coast) also reveal differences in commuting time characteristics which are consistent with Mokhtarian and Chen's claims that constant travel time hypothesis applies only at the aggregate level [25]. Given the limitations in the datasets and in our methodology, we did not give a final answer on the constant travel budget hypothesis, but leave room for thought and specific testable directions for future works. A very interesting continuation for future studies will focus on expanding the scope of the datasets to a larger number of cities and countries and testing this observation more generally with more accurate measurements (for example, supplements from other fine-grained smart phone trajectory traces). Ultimately, any observations about human mobility that apply on the universal level do not only help us gain insight into the fundamental characteristics of how we move and budget our time, but also have profound policy-level implications in urban and transportation planning.

**Acknowledgements**

The authors would like to acknowledge Orange, Octo Telematics, AirSage, and Saudi Telecom Company for providing the datasets. The authors would also like to thank Sebastian Grauwin, Michael Szell, Markus Schlapfer, and the rest of the MIT Senseable Laboratory for providing feedback on this paper. We further thank Ericsson, the MIT SMART Program, the Center for Complex Engineering Systems (CCES) at KACST and MIT CCES program, the National Science Foundation, the MIT Portugal Program, the AT&T Foundation, Audi Volkswagen, BBVA, The Coca Cola Company, Expo 2015, Ferrovial, The Regional Municipality of Wood Buffalo and all the members of the MIT Senseable City Lab Consortium for their multi-faceted support of the research.

Exploring universal patterns in human home-work commuting from mobile phone data

Exploring universal patterns in human home-work commuting from mobile phone data

**Tables**

Table 1. Summary of Datasets. The fraction represented shows the approximated ratio of the number

Exploring universal patterns in human home-work commuting from mobile phone data of users tracked by each dataset to the total number of potential users in the regions concerned.

| # | Name | Sponsor | Nature | Scope | Year | Fraction represented |
|---|---|---|---|---|---|---|
| 1 | "Ivory Coast" | Orange | Mobile phone signaling data | Country | 2011-2 | 2.5% |
| 2 | "Portugal" | Orange | Mobile phone signaling data | Country | 2006-7 | 19% |
| 3 | "Saudi Arabia" | STC | Mobile phone signaling data | Country | 2012-13 | 49% |
| 3 | "Boston" | AirSage | Mobile phone signaling data | City | 2009 | 43% |
| 4 | "Milan" | Octo Telematics | GPS traces | City | 2013 | 7.5% |

Table 2. Statistics from Spearman's correlation tests on the timing of commute, with data drawn from Fig. 4.

|  | Morning commute | | Evening commute | |
|---|---|---|---|---|
|  | Correlation coefficient ($\rho$) | Significance level (p-value) | Correlation coefficient ($\rho$) | Significance level (p-value) |
| Portugal (median time) | -0.9166 | 0.0101 | 0.9822 | 0.0005 |
| Ivory Coast (median time) | -0.9260 | 0.0008 | -0.7773 | 0.0689 |
| Portugal (Gaussian mean time) | -0.9644 | 0.0080 | -0.9378 | 0.0185 |
| Ivory Coast (Gaussian mean time) | -0.9399 | 0.0175 | 0.7996 | 0.1044 |

**Figure Legends**

Figure 1. Range of mobility during day and night. This is quantified by the mean daily dwell-time that an average individual in Ivory Coast (right plot) and Portugal (left plot) spends in each of his/her ranked places in the set of non-overlapping frequented places, plotted on a log-log scale, during the day (solid red lines with crosses) and during the night (dashed blue lines with open circles). While the daytime curves follow roughly Zipf's law, the nighttime curves show a distinct sigmoidal behavior.

Figure 2. Distributions of home-work commuting distances, aggregated by countries/cities. The distributions are plotted for Ivory Coast (blue solid line with closed dots), Portugal (red dashed line with x's), Saudi Arabia (green solid line with open circles), Boston (black solid line with open diamonds), and Milan (cyan dashed line with open triangles). The inset plot is the same plot, reproduced on a log-log scale to show long-tail behaviors of the distributions. The same plots, as cumulative density functions, are shown in Figure S4 for comparative purposes.

Figure 3. Distributions of commute timing. The timing of morning (a, c) and evening commutes (b, d) for Ivory Coast (a, b) and Portugal (c, d), for individuals binned by their commute distances: < 2.5 km (blue solid line), 2.5-5 km (red dashed line), 5-10 km (green dash-dotted line), 10-20 km (black solid line), and > 20 km (cyan dashed line). The individual's commute times in the morning and evening are estimated, respectively, by the time of the last call from home in the morning, and by the time of the first call from home in the evening. Fig. S3 shows the sample fit of such distribution to a Gaussian distribution.

Figure 4. Peak commute times as a function of commute distance. The peak times of morning (a, c) and evening (b, d) commutes for Ivory Coast (blue solid line with open circles) and Portugal (red dashed line with closed dots), as a function of commute distance. There are two methods of calculating this peak time: the median time (a, b), and the fitted Gaussian mean time (c, d). Note the stronger distance-dependent behaviors in the morning. The individual's commute times in the morning and evening are estimated, respectively, by the time of the last call from home in the

Exploring universal patterns in human home-work commuting from mobile phone data morning, and by the time of the first call from home in the evening. Fig. S3 shows the procedure whereby Gaussian distributions are fitted to the distributions plotted in Fig. 3 in order to produce the peak commute time values. The statistics from the Spearman's rank correlation tests on these relationships are summarized in Table 2.

Figure 5. Mean commute times as a function of commute distance. The means are for the morning (a) and evening (b) for Ivory Coast (blue solid line with closed dots), Portugal (red dashed line with x's), Saudi Arabia (green solid line with open circles), Boston (black solid line with open diamonds), and Milan (cyan dashed line with open triangles). While Ivory Coast, Portugal, and Boston consist of mobile phone datasets that cover aggregate commute patterns, the Milan data are GPS traces, which provide a comparative insight into car-only commute patterns.

Figure 6. Probability density functions of commute times. The first column of figures shows the probability density functions of morning commute times based on mobile phone signaling data, in Ivory Coast (a-b), Portugal (c-d), Saudi Arabia (e-f), Boston (g-h), and Milan (i-j), for individuals binned by their commute distances: < 5 km (blue solid line), 5-10 km (red dashed line), 10-20 km (green dash-dotted line), 20-40 km (black solid line), and 40-80 km (magenta dashed line). The inset plots show the cumulative distribution function of the same quantities. The second column of figures shows the probability density functions of evening commute times the respective regions.

**Supplementary Figure Legends**

**Figure S1.** The mean commute times and populations of selected major American cities. The commute times (blue bars) and populations (cyan line) are from the 2010 American Community Survey[16]. The cities, ranked by their sizes, include, from left to right: New York, Los Angeles, Chicago, Dallas, Houston, Philadelphia, Washington, Miami, Atlanta, Boston, San Francisco, Detroit, San Bernardino, Phoenix, Seattle, Minneapolis, San Diego, St. Louis, Tampa, Baltimore, Denver, Pittsburgh, Portland, Sacramento, San Antonio, Orlando, Cincinnati, Cleveland, and Kansas City.

**Figure S2.** Sample data visualization. (a) The accumulated density plot of GPS positions in the Milan Metropolitan Area, with red showing locations with the highest frequency of GPS position reporting, and blue showing locations with the lowest frequency of GPS position reporting, over a period of one day. (b) The mobility pattern of an individual over 7 days in Milan, with red dots showing the original reported GPS positions, and blue open circles showing locations where the individual spends the most time. Based on the day-time and night-time activities, the home and work locations are identified and labeled. The values on the *x* and *y* axes (longitudes and latitudes) have been removed from this plot in order to protect the user's anonymity. (c) Map showing the density plot of mobile phone towers in the Saudi Arabia dataset.

**Figure S3.** (a) A Gaussian fit of the distribution of peak commute time, for the example of morning commuters in Portugal. (b) The corresponding Q-Q plot of the same Gaussian fit, showing that for most of the commuting time domain (between 5 a.m. and 10 a.m.), the Gaussian fit is a reasonable fit.

**Figure S4.** Cumulative distribution function of commuting distances in Ivory Coast (blue solid line with closed dots), Portugal (red dashed line with x's), Saudi Arabia (green solid line with open circles), Boston (black solid line with open diamonds), and Milan (cyan dashed line with open triangles).

Exploring universal patterns in human home-work commuting from mobile phone data

**List of Figures**

Figure 1.

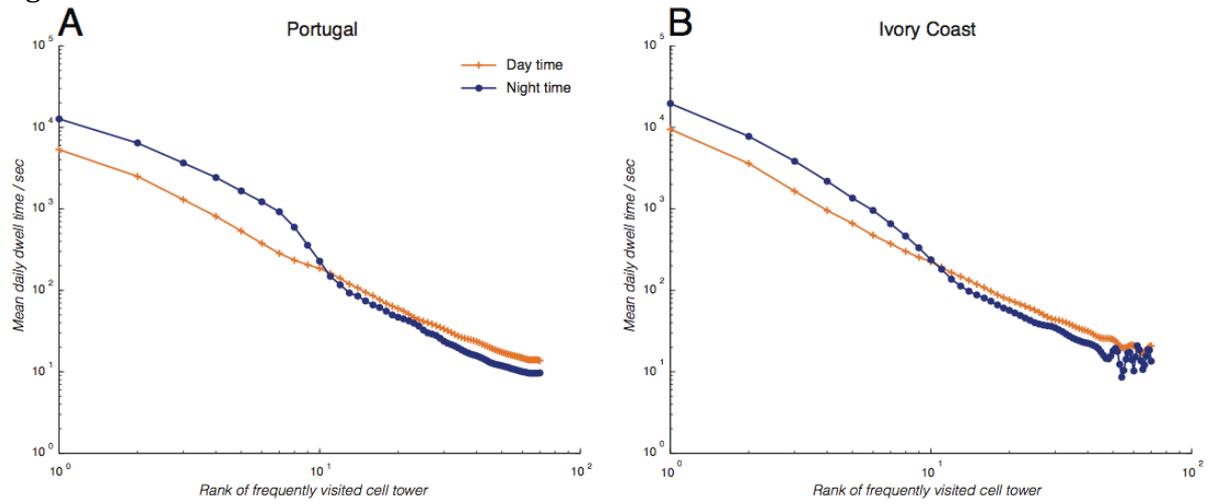

Figure 2.

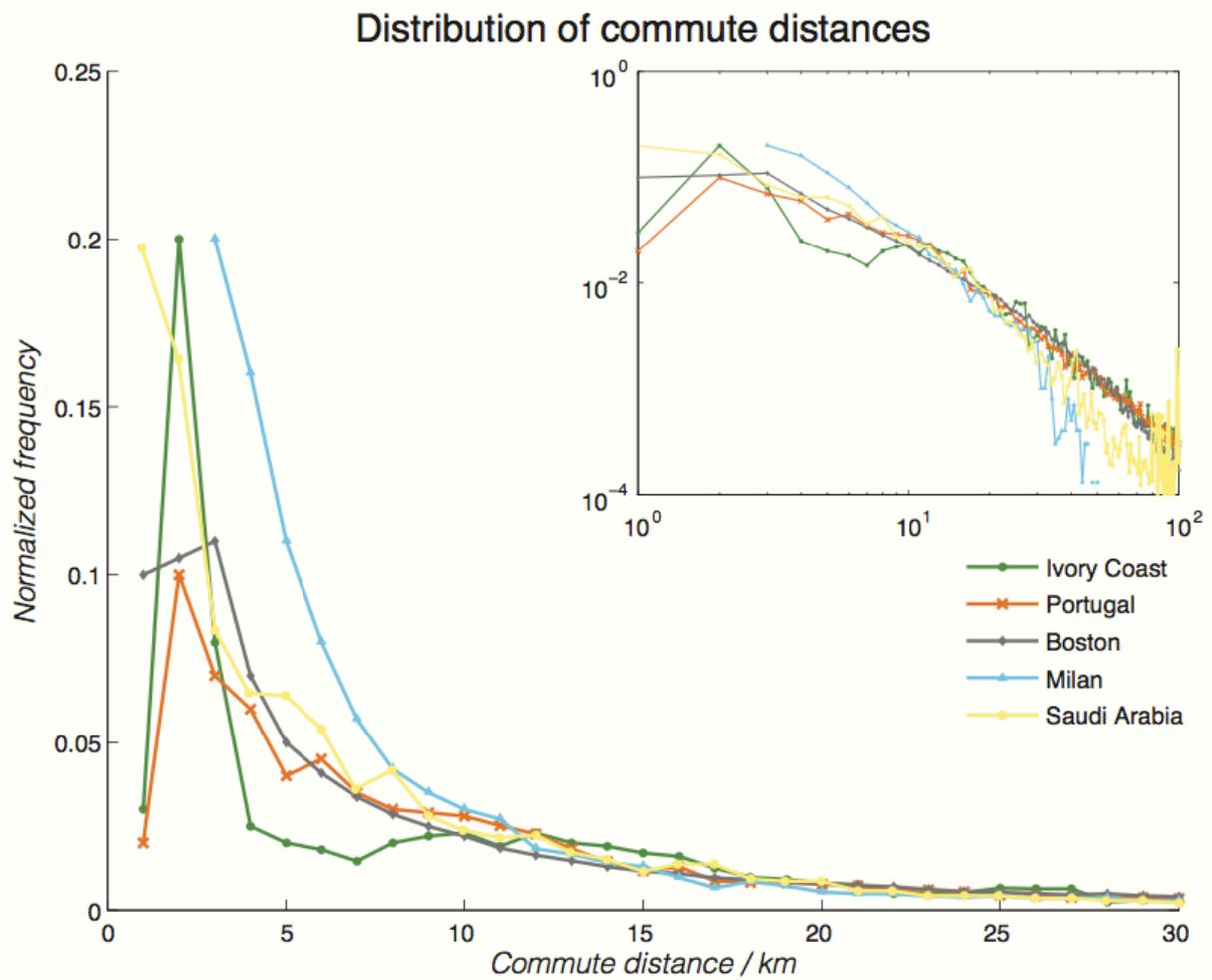

Exploring universal patterns in human home-work commuting from mobile phone data

Figure 3.

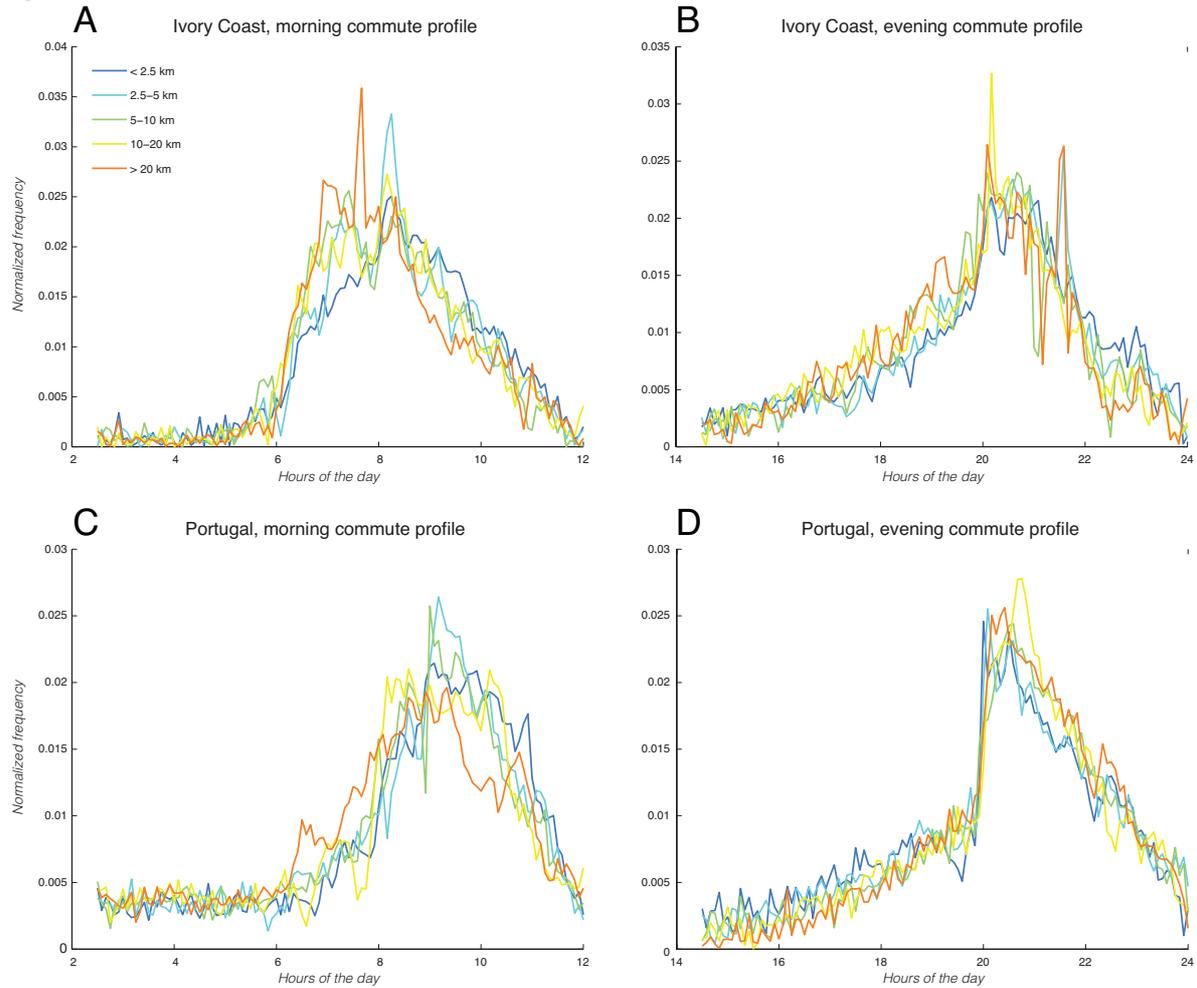

Figure 4.

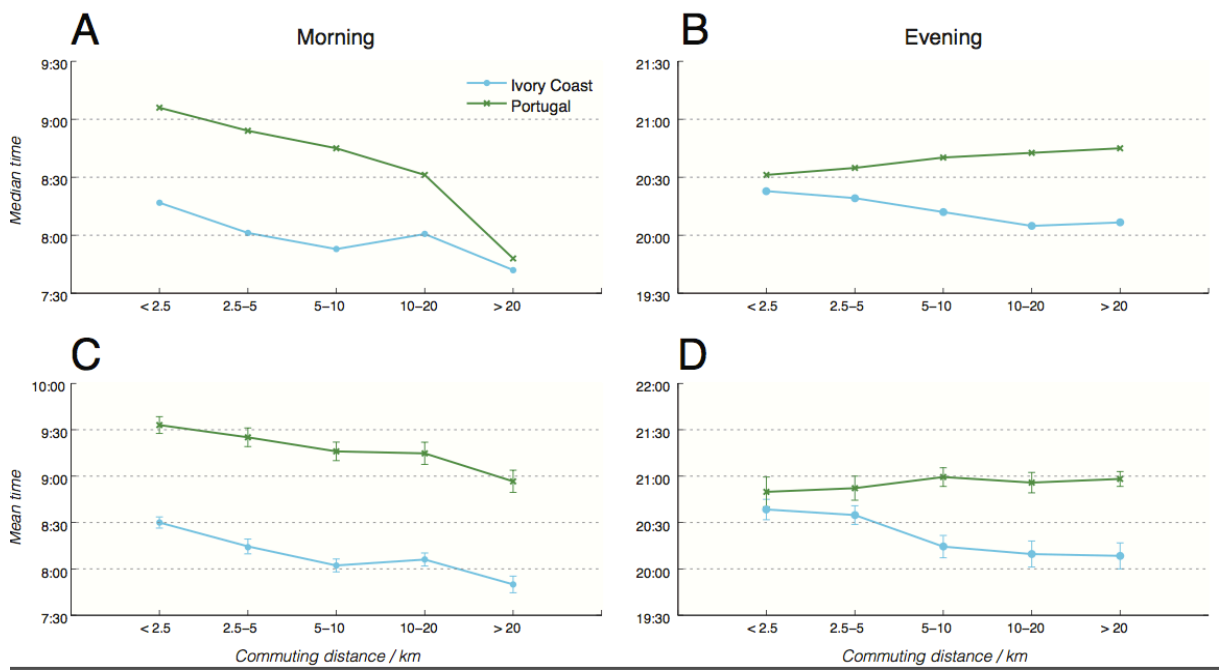

Exploring universal patterns in human home-work commuting from mobile phone data

Figure 5.

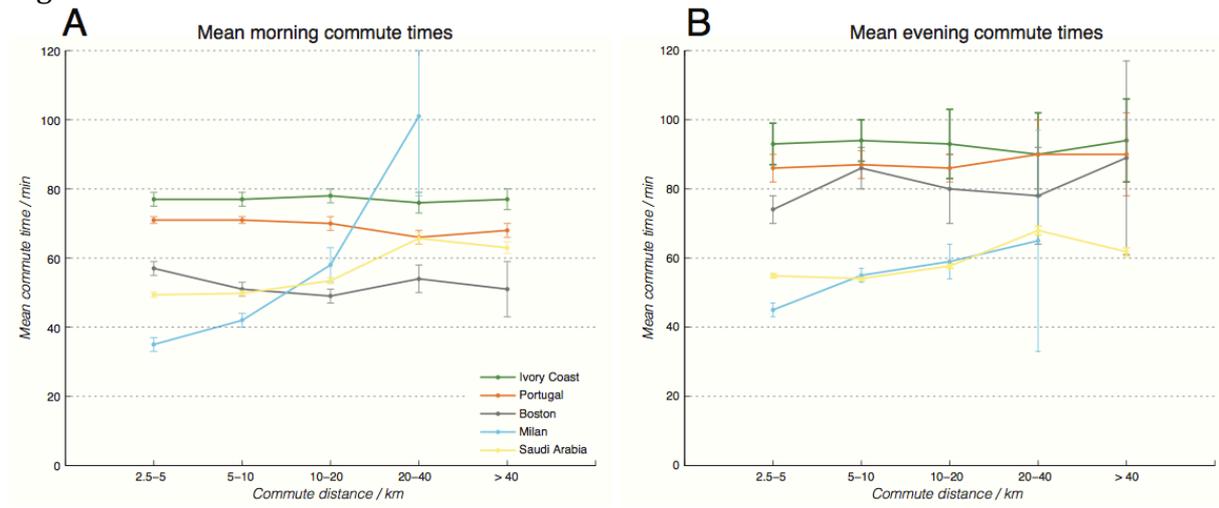

Exploring universal patterns in human home-work commuting from mobile phone data

Figure 6.

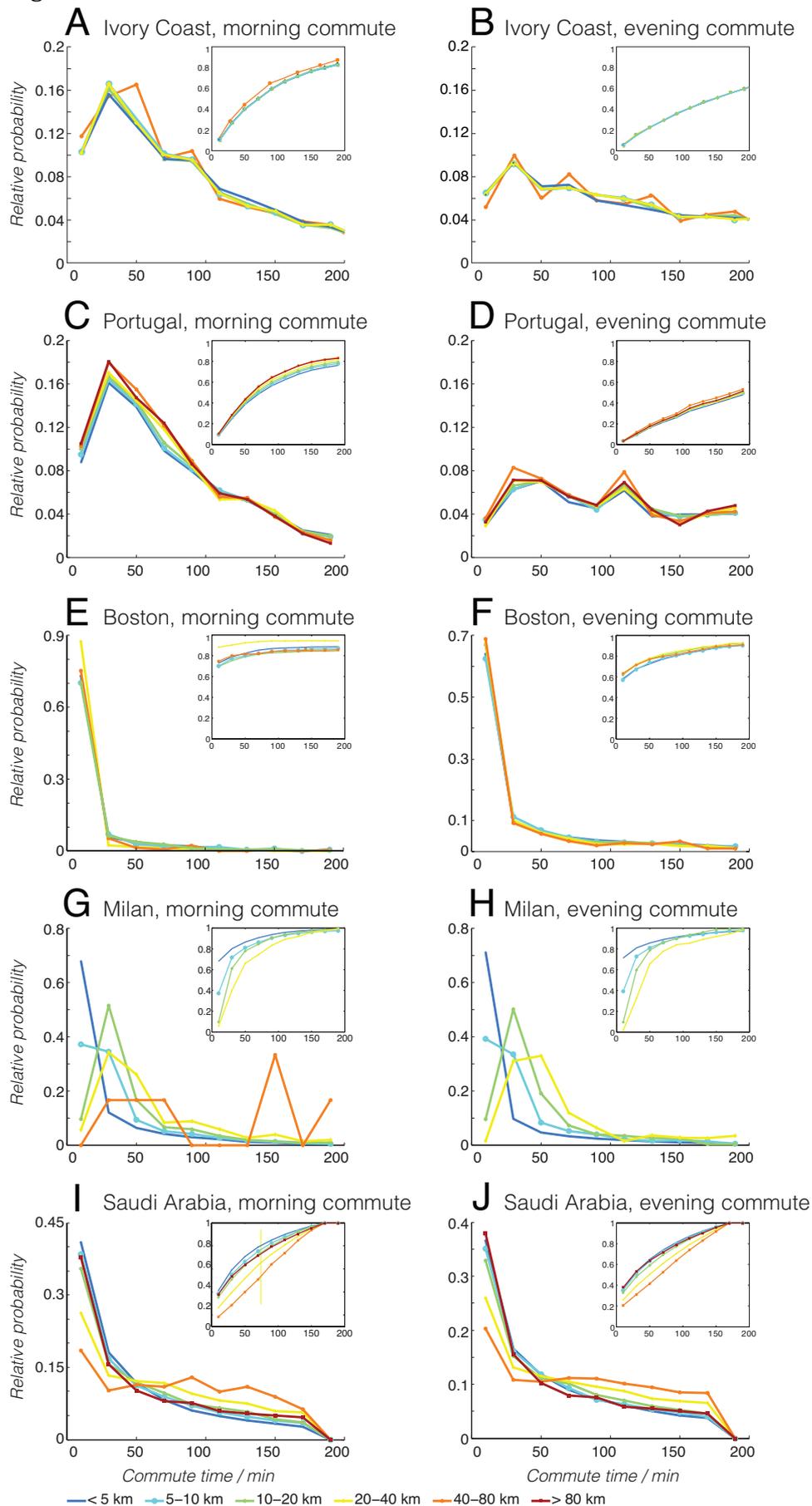

Exploring universal patterns in human home-work commuting from mobile phone data

Figure S1.

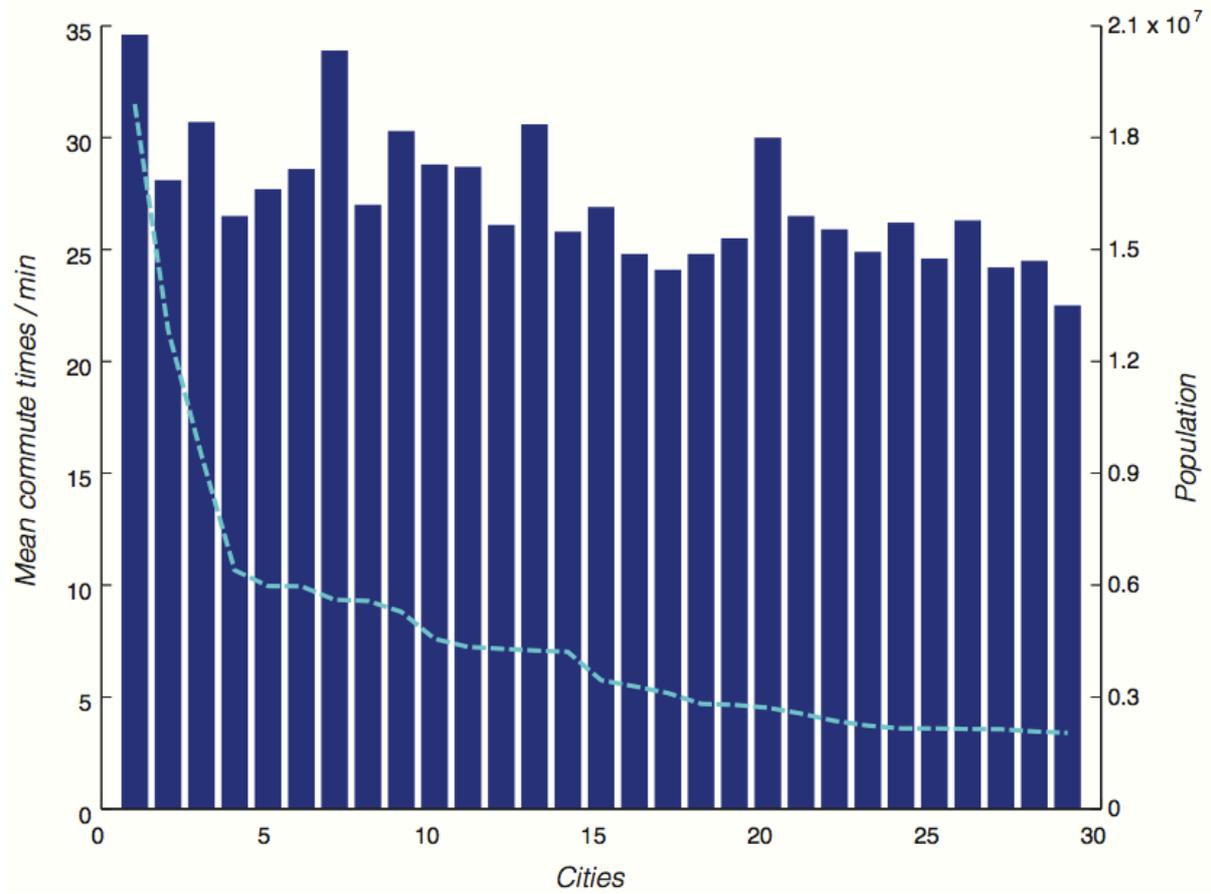

Figure S2.

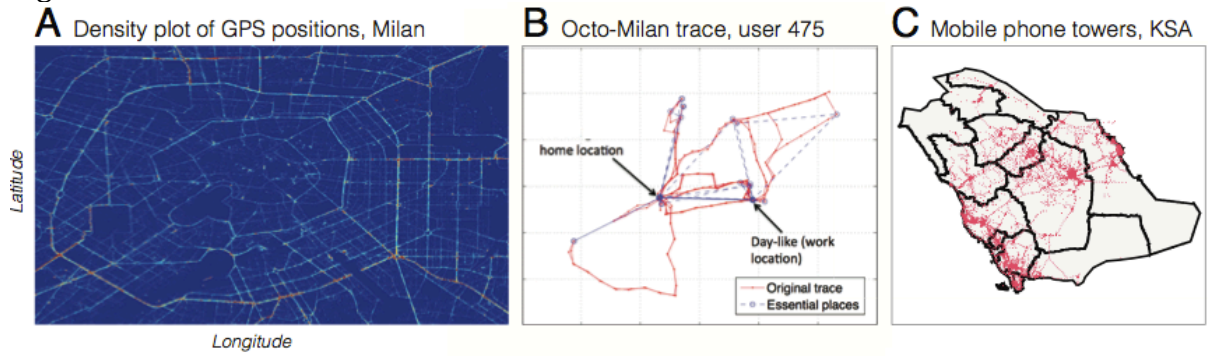

Exploring universal patterns in human home-work commuting from mobile phone data

Figure S3.

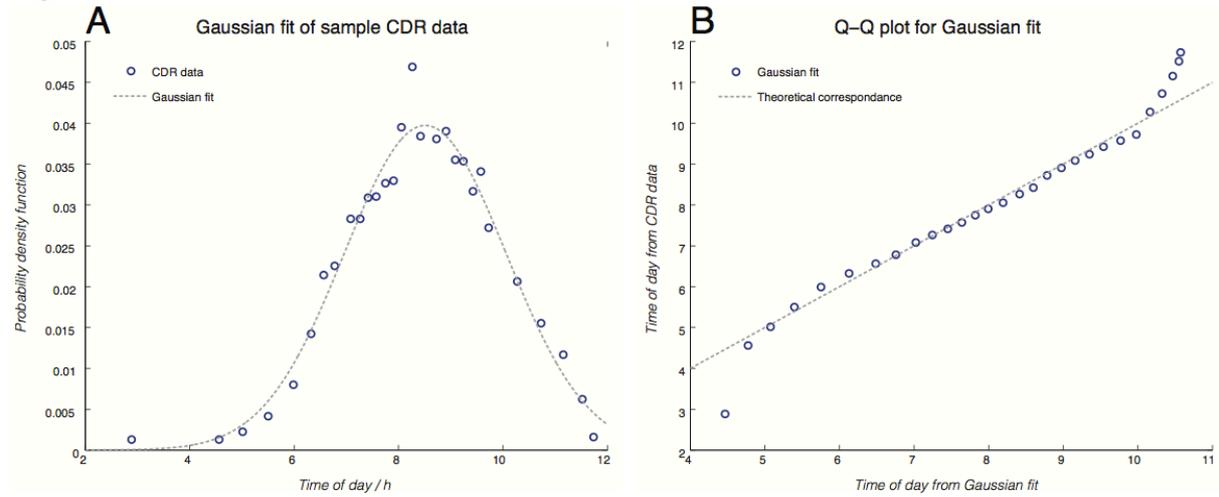

Figure S4.

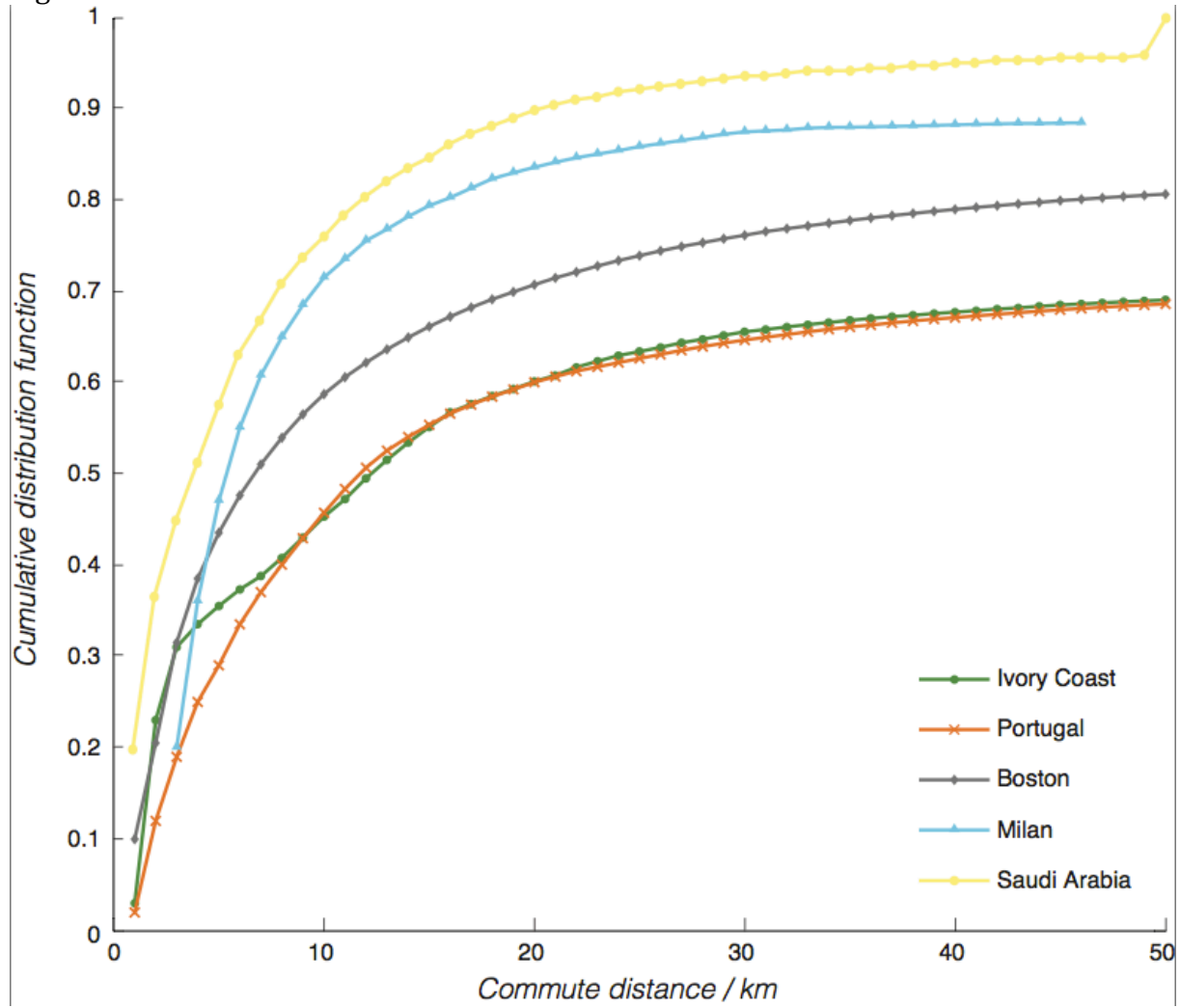